\documentclass[aps,notitlepage]{revtex4-1}
\usepackage{graphicx}
\usepackage{amsmath}
\usepackage[colorlinks]{hyperref}
\usepackage{setspace}

\begin{document}

% Title portion
\title{Neutralino Pair Production at the Photon-Photon Collider for the $\tilde{\tau}$-Coannihilation Scenario}

\author{Nasuf SONMEZ}
\affiliation{Ege University, Izmir, Turkey}

\begin{abstract}
Supersymmetry (SUSY) is a theory that gives an explanation for the strong and electroweak interactions from the grand unification scale down to the weak scale. 
The search for supersymmetric particles still continues at full speed at the LHC without success. 
The main tasks at the ILC is complementing the LHC result and also search for new physics.
In this study, the neutralino pair production via photon-photon collision is studied for the $\tilde{t}$-coannihilation scenario in the context of MSSM at the ILC. 
In the calculation, all the possible one-loop diagrams are taken into account for the photon-photon interaction. 
We present the production cross section and distribution of various observables for the lightest and next-to-lightest neutralino pairs for benchmark models which are specifically presented in the light of LHC8 data analysis, employing these benchmark models for neutralino pair production could show the potential of the ILC concerning the dark matter searches in supersymmetry.
\end{abstract}

\maketitle

%%%%%%%%%%%%%%%%%%%%%%%%%%%%%%%%%%%%%%%%%%%%
%% MAINMATTER
%%%%%%%%%%%%%%%%%%%%%%%%%%%%%%%%%%%%%%%%%%%%

\setstretch{1.5}

%%%%%%%%%%%%%%%%%%%%%%%%%%%%%%%%%%%%%%%%%%%%
\section{INTRODUCTION}
Despite the success of the Standard Model (SM), the scalar sector still suffers from the quadratic divergence. 
One possible solution to that is to introduce a new kind of symmetry, which handles these divergences and reduces to merely logarithmic.
This theory is called supersymmetry and it 
%Supersymmetry is a theory which 
gives an explanation for the strong and electroweak interactions at the Planck scale down to the weak scale. % by introducing supersymmetric partners for all the SM particles.
According to the theory, if R-parity is conserved in the decay of the supersymmetric particles, the lightest supersymmetric particle becomes a typical candidate for the weakly-interacting dark matter\cite{Martin:1997ns}.
The lightest supersymmetric particle in the theory is called neutralino ($\tilde\chi_1^0$) and due to the weakly interacting nature it will escape the detector without leaving any signal.
However, it could be tracked via the missing energy in each event.

%THE CURRENT SITUATION ON SUSY SEARCH
The LHC collider at CERN provided pp collisions at $\sqrt{s}=7\;  {TeV}$ (LHC7) and $\sqrt{s}=8\;  {TeV}$ (LHC8) in Run-1
%with $5\;fb^{-1}$ and $20\;fb^{-1}$ total luminosity respectively, strong limits on SUSY parameter space and constraints on the mass of the sparticles are set \cite{Khachatryan:2014mma,Khachatryan:2014qwa,Khachatryan:2014doa}.
%Using LHC8 and LHC7 
to the ATLAS and CMS experiments. 
%After the analysis campaign of the collision data,
These experiments excluded a very favorable model known as CMSSM for the following cases \cite{CMS:zxa}; i.) $m_{\tilde{g}} \leq1500\;  {GeV}$ and $m_{\tilde{g}}\simeq m_{\tilde{q}}$, and ii.) $m_{\tilde{g}} \leq1000\;  {GeV}$ and $m_{\tilde{g}}\ll m_{\tilde{q}}$.
In addition to that, if the discovered Higgs boson emerges to be the supersymmetric light CP-even Higgs boson, gauge and anomaly mediated supersymmetry breaking models will also be ruled out.

%WHAT I DID IN THE CALCULATION
In this work, the neutralino pair production including whole set of all possible one-loop level Feynman diagrams are calculated.
The numerical calculation is presented for the \emph{s-tau coannihilation} scenario (STC) which is proposed in the light of LHC7 and LHC8 results \cite{Baer:2013ula}.
The masses of neutralinos and charginos in this benchmark point are at the sub-TeV range, which makes them accessible at the ILC in $\gamma\gamma$ collision mode.
Therefore, this benchmark point is outside of the limits presented by the whole LHC program and it could be reached at the ILC, as it is concluded in this work.
The total cross section as a function of center-of-mass (CM) energy for the neutralino pairs and the angular distribution of the cross section are calculated for the unpolarized photon-collisions.
In addition to these, the total integrated cross section is calculated for $\sqrt{s}=0.5-1\; {TeV}$ CM energies. 

%PREVIOUS  PAPERS AND RESULTS
%Neutralino pair production rates in $\gamma\gamma$ collider, especially, helicity nature of the cross section are previously studied by G.J.Gounaris et.all. \cite{Gounaris:2003ti}.

%%%%%%%%%%%%%%%%%%%%%%%%%%%%%%%%%%%%%%%%%%%%
\section{THEORETICAL FORMULATION}
The process for the neutralino pair production at the NLO via photon-photon collision is denoted as 
\begin{eqnarray*}
\gamma (k_1,\mu)\;\gamma(k_2,\nu)\;\rightarrow \;  \tilde{\chi}^0_i (k_3)\;\tilde{\chi}^0_j(k_4) \;\;\; (i,j=1,2)\,,\nonumber
\end{eqnarray*}
where $k_a$ $(a=1,...,4)$ are the four-momenta of the incoming photons and outgoing neutralinos, respectively.
$\mu$ and $\nu$ are the polarization vectors of the incoming photons. 
Since photon doesn't couple to itself, the neutralino pair production via photon-photon collision is possible via loops at the lowest.
All the Feynman diagrams contributing to the process $\gamma \gamma \rightarrow  \tilde{\chi}^0_i \tilde{\chi}^0_j$ at the one-loop level are given in \cite{Sonmez:2014ika}.
The numerical evaluation of the process $(\gamma\gamma\rightarrow \tilde{\chi}_i^0\tilde{\chi}_j^0)$ have been performed using the following packages;
  \texttt{FeynArts}\cite{Kublbeck:1992mt,Hahn:2000kx} to generate the Feynman diagrams and the corresponding amplitudes, 
  \texttt{FormCalc}\cite{Hahn:2006qw} to simplify the fermion chains then square the corresponding amplitudes and   
  \texttt{LoopTools}\cite{Hahn:1998yk} to evaluate the scalar and tensor one-loop integrals.  
There is no ultraviolet divergence in the results, because all the possible one loop level diagrams are taken into account.

The corresponding Lorentz invariant matrix element for the one-loop level process is written as a sum over box-dagrams, triangle-diagrams, bubble s-channel type and quartic ones are showed in \cite{Sonmez:2014ika},
\begin{equation}
{\cal M}= {\cal M}_{box}+ {\cal M}_{tri}+ {\cal M}_{quart}\,.
\end{equation}

In the numerical calculation, the scattering amplitude is evaluated in the CM frame, 
then the polarization vectors of the incoming photons and the helicities of the neutralinos are summed.
As a result, the cross section of the unpolarized photon collisions is given as 
\begin{equation}
\hat{\sigma}_{\gamma\gamma\rightarrow \tilde\chi_i^0\tilde\chi_j^0}(\hat{s})=\frac{ \lambda( \hat{s},m_{\tilde\chi_i^0}^2, m_{\tilde\chi_j^0}^2 )}{16 \pi \hat{s}^2} \left(\frac{1}{2}\right)^{\delta_{ij}}\frac{1}{4} \sum_{hel}{|\mathcal{M}|^2}\,,
\label{eq:partcross}
\end{equation}
where 
$\lambda( \hat{s},m_{\tilde\chi_i^0}^2, m_{\tilde\chi_j^0}^2 )$ is calculated for the outgoing neutralino pairs.

The photon beam at the ILC is formed using the laser back-scattering technique on electron beam in $e^+e^-$ collisions.
Therefore, the big fraction of the CM energy of the electron beam could be transferred to the photon collisions.
Then, the $\gamma\gamma\rightarrow \tilde\chi_i^0\tilde\chi_j^0$ process could be taken as a subprocess in the $e^+e^-$ collisions. 
Thus, the total cross section of $e^+e^-\rightarrow \tilde\chi_i^0\tilde\chi_j^0$ could easily be calculated by convoluting the photonic cross section $\hat{\sigma}_{\gamma\gamma\rightarrow \tilde\chi_i^0\tilde\chi_j^0}(\hat{s})$ with the photon luminosity in $e^+e^-$ collider.
The total integrated photonic cross section is defined in the following equation
\begin{equation}
\sigma(s)=\int_{x_{min}}^{x_{max}} \hat{\sigma}_{\gamma\gamma\rightarrow  \tilde\chi_i^0\tilde\chi_j^0}( \hat{s};\; \hat{s}=z^2s ) \frac{dL_{\gamma\gamma}}{dz}\;dz\,,
\label{eq:foldcross}
\end{equation}
where $s$ and $\hat{s}$ are the CM energy in $e^+e^-$ collisions and $\gamma\gamma$ subprocess, respectively. 
$x_{min}$ is the threshold energy for neutralino pair and defined as $x_{min}=(m_{\tilde\chi_i^0}+m_{\tilde\chi_i^0})/\sqrt{s}$.
The maximum fraction of the photon energy is taken as $x_{max}=0.83$ and the distribution function of the photon luminosity is given in \cite{Telnov:1989sd}.

%%%%%%%%%%%%%%%%%%%%%%%%%%%%%%%%%%%%%%%%%%%%
%%%%%%%%%%%%%%%%%%%%%%%%%%%%%%%%%%%%%%%%%%%%
\section{NUMERICAL RESULTS AND DISCUSSION}

In this section, the total cross section as a function of CM energy, the angular distribution of the total cross section, and the total integrated photonic cross section for the $\tilde{\tau}$-coannihilation benchmark point are presented.
The total integrated photonic cross section for each neutralino pairs are presented in Table-\ref{tab:a} for two distinct CM energies ($\sqrt{s}=0.5\,{TeV}-1.0\,{TeV}$).

In many constrained susy models, the sfermion masses are correlated at a high energy scale and light sleptons with the mass of $\sim 100-200 \;GeV$ are not allowed.
However, in a model like pMSSM the spectra of the light sleptons and heavy squarks are possible due to the higher parameter freedom.
In the pMSSM, the following parameters are taken \cite{Baer:2013ula};
\begin{itemize}
\item Higgs sector parameters : $\tan\beta=10$, $\mu=m_A=400 \;GeV $
\item Trilinear couplings : $A_t=A_b=A_\tau=-2.1\;GeV$
\item Gaugino mass parameters : $M_3=2\; TeV$, $M_2=210\; GeV$
\item Slepton mass parameters : $m_L(1,2,3)=205\; GeV$, $m_E(1,2,3)=117.5\; GeV$
\item Squark mass parameters : $m_Q(1,2)=m_D(1,2)=m_U(1,2)=2\; TeV$, $m_Q(3)=1.5\;TeV$, $m_U(3)=500\;GeV$, $m_D(3)=800\;GeV$
\end{itemize}
The SM parameters and the electroweak scale parameter set of the benchmark point are adopted from \cite{bib:webpage} and Ref. \cite{Eidelman:2004wy}, respectively.
More information for the benchmark point could be found at \cite{Baer:2013ula} and references therein. 
At this benchmark point the neutralino masses are $m_{\tilde{\chi}_{1,2}^0}\approx (96, 206) \;{GeV}$, the charginos have mass ($m_{\tilde{\chi}_{1,2}^{\pm}}\approx (206, 425) \;{GeV}$), and stau masses are $m_{\tilde{\tau}_{1,2}}\approx (107, 219) \;{GeV}$.
The mass spectrum of the neutral Higgs bosons in STC are $m_{h_0/H_0/A_0}=(124,401,400)\,{GeV}$.
The favor of this benchmark point is that it offers an adequate dark matter annihilation mechanism due to the light $\tilde{\tau}_1$ mass.

\begin{figure}
	\includegraphics[height=6cm]{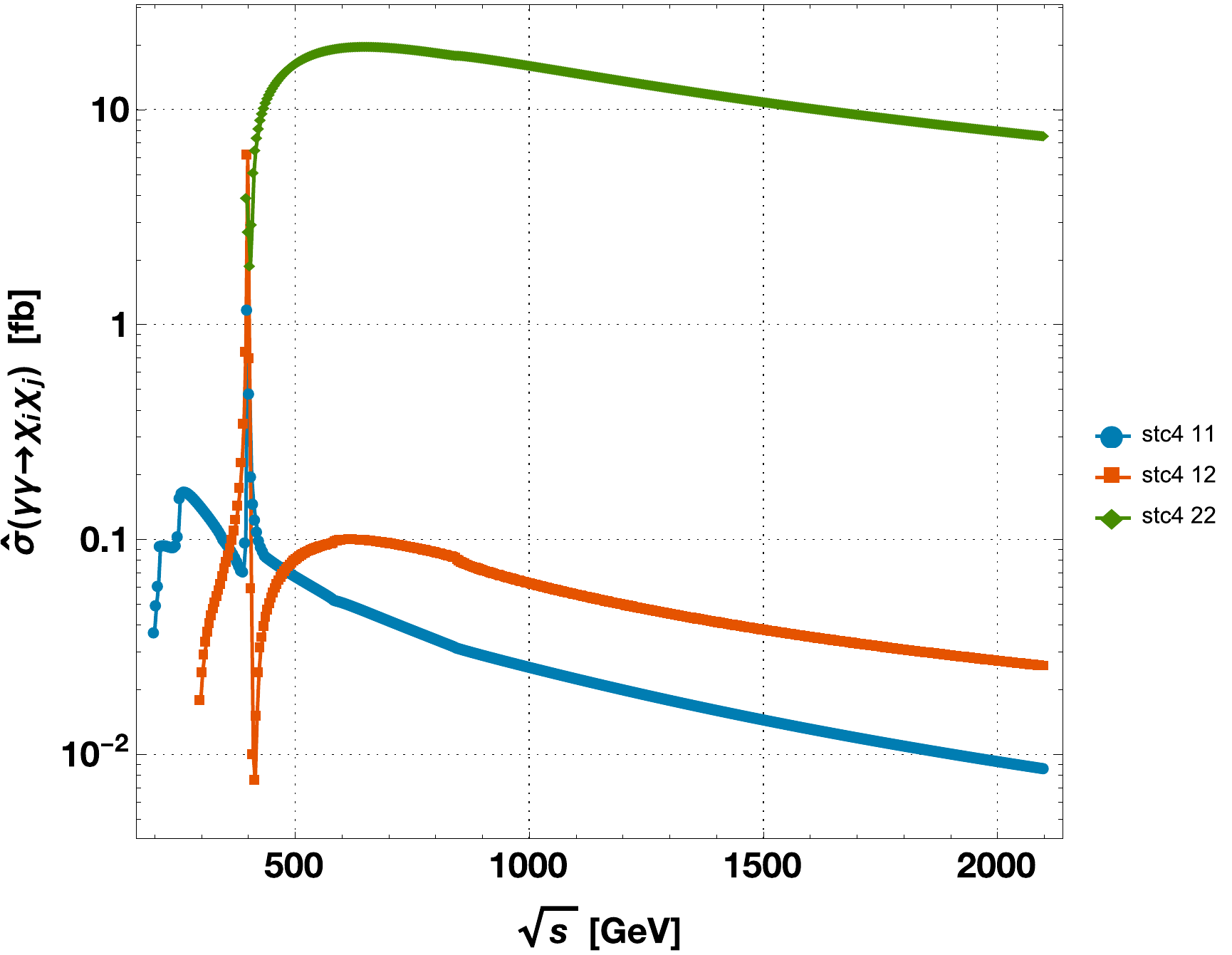}
\caption{The total cross section as a function of CM energy for the $\tilde{\chi}^0_1\tilde{\chi}^0_1$,  $\tilde{\chi}^0_1\tilde{\chi}^0_2$ and $\tilde{\chi}^0_2\tilde{\chi}^0_2$ pairs.}
\label{fig:fig1}
\end{figure}
After plugging in all the relevant parameters into the expression, the numerical results for the cross section are obtained. 
The energy dependence of the $\tilde{\chi}^0_1\tilde{\chi}^0_1$,  $\tilde{\chi}^0_1\tilde{\chi}^0_2$ and $\tilde{\chi}^0_2\tilde{\chi}^0_2$ pairs in unpolarized photon collisions are given in Figure \ref{fig:fig1}.
It can be seen that the cross section goes up to $17\,{fb}$ around $\sqrt{\hat s}=650\,{GeV}$ for $\tilde{\chi}^0_2\tilde{\chi}^0_2$ pair.
Therefore, the production rates are significantly lower around $0.1\,{fb}$ for the $\tilde{\chi}^0_1\tilde{\chi}^0_1$ and $\tilde{\chi}^0_1\tilde{\chi}^0_2$ pairs at CM energies higher than $500\; GeV$.
However, the cross section is enhanced for the $\tilde{\chi}^0_1\tilde{\chi}^0_1$ and $\tilde{\chi}^0_1\tilde{\chi}^0_2$ pairs and it gets value up to $5\,{fb}$ due to the s-channel enhancement.
When the CM energy of the incoming photons are close to the masses of the neutral $h_0$, $H_0$ and $A_0$ bosons, the s-channel resonance effect in the diagrams where neutral supersymmetric Higgs bosons are intermediated enhances the cross section significantly.
The spikes seen in Figure-\ref{fig:fig1} around $\sqrt{\hat s}=400\,{GeV}$ are due to the neutral Higgs mediation.
For the numerical analysis, the decay widths of the neutral Higgs particles are calculated at the NLO level accuracy using \texttt{FeynHiggs} \cite{Heinemeyer:1998yj,Hahn:2009zz}.
\begin{figure}
	\includegraphics[height=5cm]{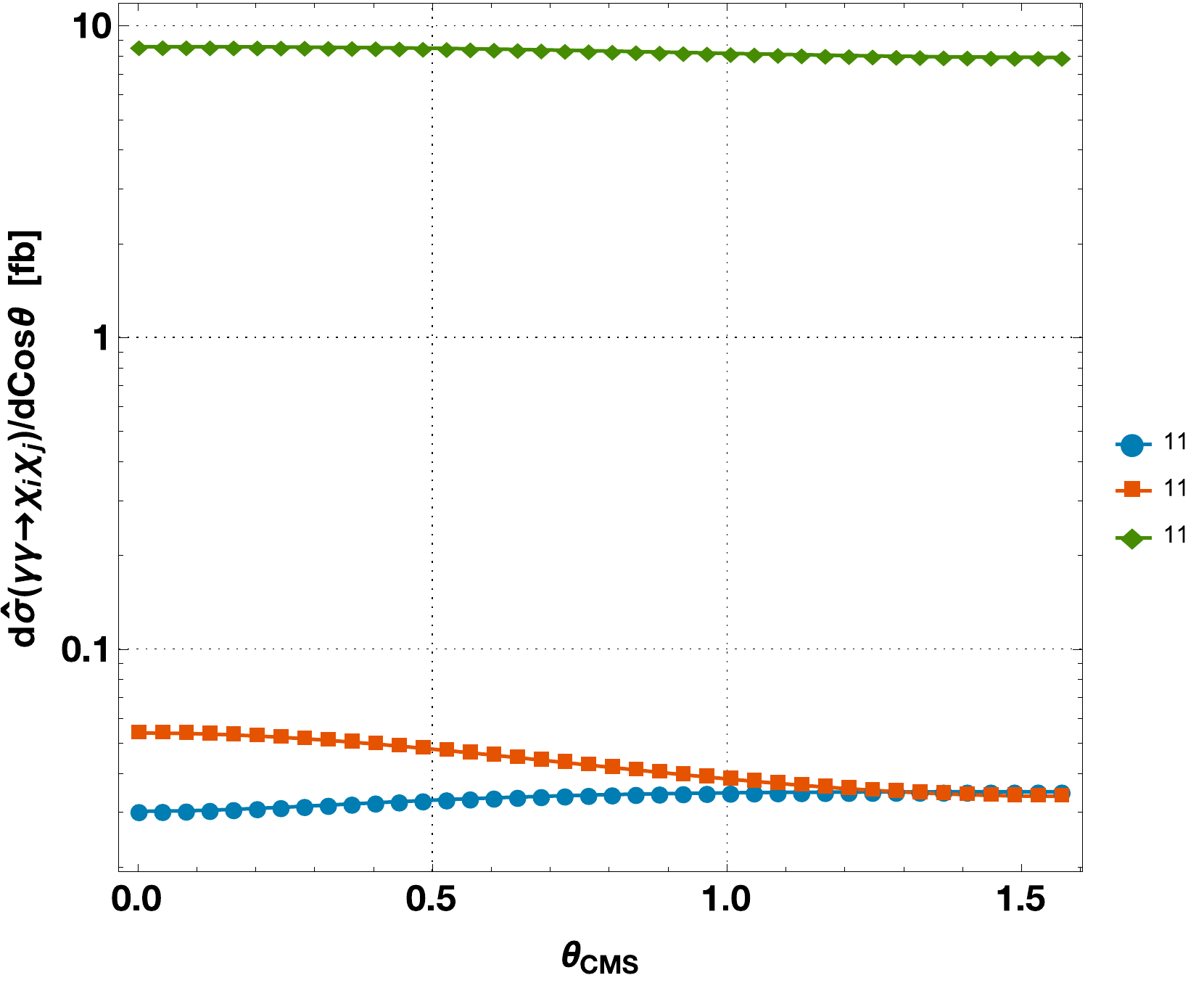}
	\includegraphics[height=5cm]{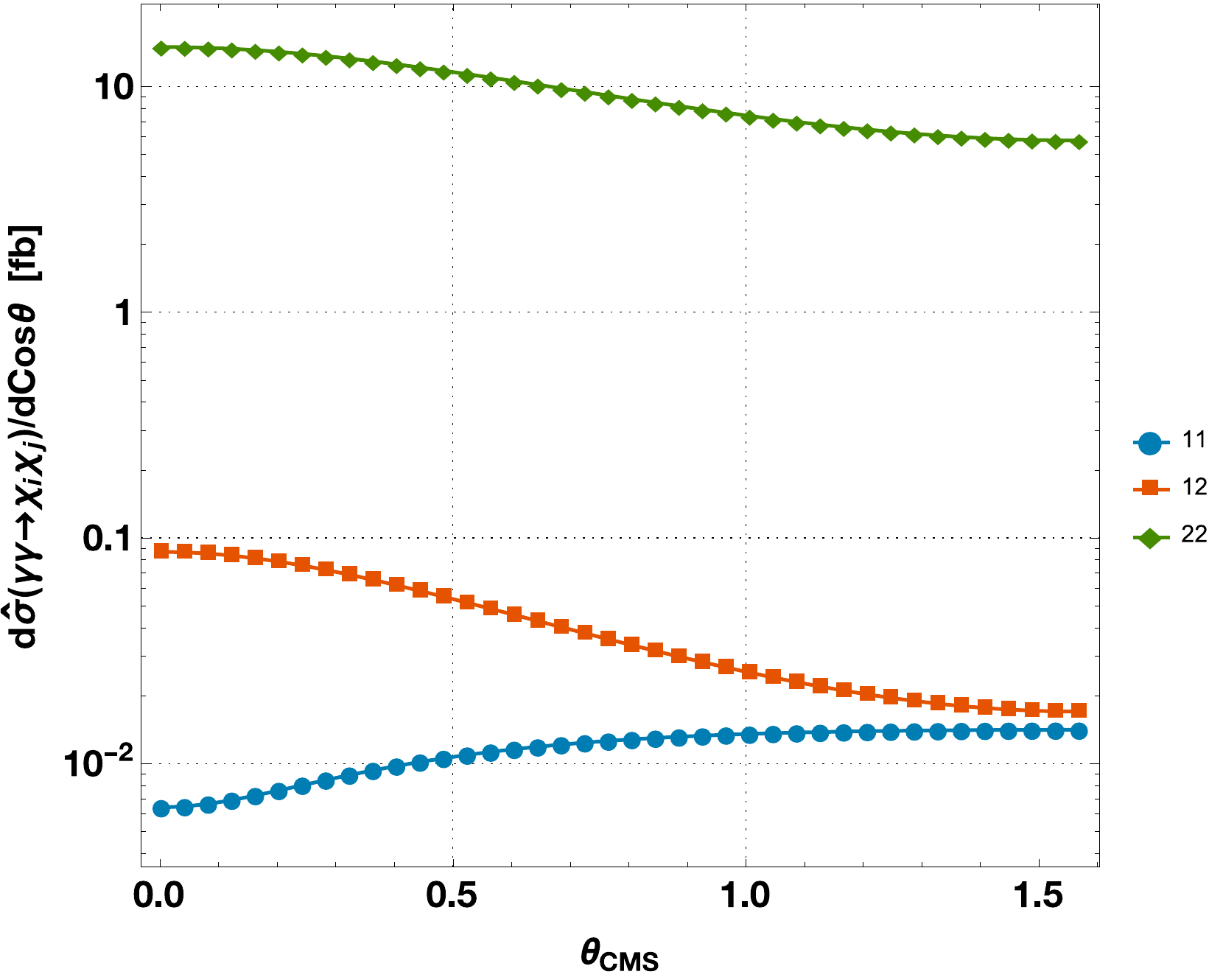}
\caption{The angular distribution of each neutralino pair for $0.5\,{TeV}$(left) and $1\,{TeV}$(right) CM energies.}
\label{fig:fig2}
\end{figure}
The angular distribution of the photonic cross section for the $\tilde{\chi}^0_1\tilde{\chi}^0_1$,  $\tilde{\chi}^0_1\tilde{\chi}^0_2$ and $\tilde{\chi}^0_2\tilde{\chi}^0_2$ pairs at two distinct CM energies $\sqrt{s}=0.5\,{TeV}-1.0\,{TeV}$ are given in Figure-\ref{fig:fig2}(left)-(right), respectively.
The angular distribution of $\tilde{\chi}^0_2\tilde{\chi}^0_2$ is close to isotropy, where $\tilde{\chi}^0_1\tilde{\chi}^0_1$ and $\tilde{\chi}^0_1\tilde{\chi}^0_2$ neutralino pairs have a small asymmetry at the $\sqrt{s}=500\;GeV$ seen in Figure \ref{fig:fig2}(left).
At the $\sqrt{s}=1\;TeV$, there is a good asymmetry for all the neutralino pairs (Figure \ref{fig:fig2}(right)).

%Therefore, that looks like where the main contribution comes to the total integrated photonic cross section calculated by Eq-\ref{eq:total_cross}, but a detailed analysis showed that the s-channel type bubble and triangle diagrams have a destructive interference and they almost cancel each other.
%In a result, box diagrams are the ones which make the dominant contribution.

\begin{table}
\begin{tabular}{l|cc}
\hline
							&	$\sqrt{s}=0.5\,{TeV}$	(fb)	&	$\sqrt{s}=1.0\,{TeV}$	(fb)\\
\hline
$\tilde{\chi}_1^0\tilde{\chi}_1^0$	&	0.06905				&	0.0744			\\
$\tilde{\chi}_1^0 \tilde{\chi}_2^0$	&	0.06616				&	0.1008			\\
$\tilde{\chi}_2^0\tilde{\chi}_2^0$	&	0.04935				&	8.875			\\
\hline
\end{tabular}
\caption{Integrated total photonic cross section of $e^+e^-\rightarrow\gamma\gamma\rightarrow\tilde\chi_i^0\tilde\chi_j^0$  for STC benchmark point at CM energy $\sqrt{s}=0.5-1.0\;TeV$.}
\label{tab:a}
\end{table}

%%%%%%%%%%%%%%%%%%%%%%%%%%%%%%%%%%%%%%%%%%%%
\section{CONCLUSION}

In this work, we have calculated the production rates of the neutralino pairs at the NLO level including all possible Feynman diagrams in a future photon collider.
The numerical analysis on production rates as a function of CM energy, angular dependence and the total integrated photonic cross section are calculated at the ILC for the STC benchmark point.
This benchmark point is reachable, particularly, at the International Linear Collider.
The production rates of $\tilde{\chi}_1\tilde{\chi}_1$,  $\tilde{\chi}_1\tilde{\chi}_2$ and $\tilde{\chi}_2\tilde{\chi}_2$ pairs via photon collisions are analyzed up to $\sqrt{\hat{s}}=2.2\,{TeV}$.
The peaks seen in the cross section distributions are due to resonance effect of the neutral Higgs propagation particularly in the triangle type diagrams.

The integrated total photonic cross section is calculated by convoluting the $\gamma\gamma \rightarrow \tilde{\chi}_i\tilde{\chi}_j$ cross section with the photon luminosities at the ILC for two distinct CM energies of the incoming $e^+e^-$ beams.
Among the results presented in \cite{Sonmez:2014ika} and the one calculated in this work, the STC gives higher total integrated cross section for the same second lightest neutralino pairs after NUHM2 benchmark point at $\sqrt{s}=1\; TeV$.
Pair production rate of the lightest neutralinos is quite lower than the other benchmark points analyzed in \cite{Sonmez:2014ika}.
However, $\tilde{\chi}_1\tilde{\chi}_2$ pair production rate gets high at $\sqrt{s}=1.0\; TeV$ CM energy.
The lightest neutralino pair without additional hard photon is not observable due to the weakly interacting nature, but the lightest and next-to-lightest neutralino pair is observable at the detector.
This process is a primary interest in the current supersymmetry searches.
For experimental point of view the lightest neutralino will produce a large missing energy in the collisions and production of missing energy with the asymmetry in the events will give a strong clue on supersymmetry in ILC.
It is concluded that a $\gamma\gamma$ collider with an additional very small cost compared to the $e^+e^-$ collider could produce new results on supersymmetry and that might give hints on the beginning of the universe.

%%%%%%%%%%%%%%%%%%%%%%%%%%%%%%%%%%%%%%%%%%%%%%%%
%% BACKMATTER
%%%%%%%%%%%%%%%%%%%%%%%%%%%%%%%%%%%%%%%%%%%%%%%%

% Acknowledgement
\section{ACKNOWLEDGEMENTS}
Author thanks for having access to the computing resources of \texttt{fencluster} at Faculty of Science in Ege University.

% References

\nocite{*}
\bibliographystyle{aipnum-cp}%
%\bibliography{sample}%

\end{document}